\documentclass{fmj2010}
\usepackage{graphicx}
\usepackage{amsmath}

\def\fgrst{\textit{Fermi Gamma-Ray Space Telescope}}
\def\fermilat{\textit{Fermi}/LAT}
\def\fermi{\textit{Fermi}}

\begin{document}
   \title{VLBI Monitoring of 3C~84 in Gamma-ray Active Phase}

   \author{H. Nagai\inst{1}, K. Suzuki\inst{2}, K. Asada\inst{3}, M. Kino\inst{4}, S. Kameno\inst{5}, A. Doi\inst{1}, M. Inoue\inst{3}, \and U. Bach\inst{6}
          }

   \institute{1. Institute of Space and Astronautical Science, Japan Aerospace Exploration Agency, 3-1-1 Yoshinodai, Chuo-ku, Sagamihara 252-5210, Japan \\ 2. Institute of Astronomy, University of Tokyo, 2-21-1 Osawa, Mitaka, Tokyo 181-0015, Japan \\ 3. Institute of Astronomy and Astrophysics, Academia Sinica, P.O. Box 23-141, Taipei 10617, Taiwan, R.O.C. \\ 4. National Astronomical Observatory, 2-21-1 Osawa, Mitaka, Tokyo 181-8588, Japan \\ 5. Faculty of Science, Kagoshima University, 1-21-35 Korimoto, Kagoshima, Kagoshima 890-0065, Japan \\ 6. Max-Planck-Institute f\"{u}r Radioastronomie, Auf dem H\"{u}gel 69, 53121 Bonn, Germany
             }

   \abstract{We report multi-epoch Very Long Baseline Interferometry (VLBI) study of the innermost jet of 3C~84.  We carried out 14-epoch VLBI observations during 2006-2009 with the Japanese VLBI Network (JVN) and the VLBI Exploration of Radio Astrometry (VERA), immediately following the radio outburst that began in 2005.  Comparison between VLBI lightcurve and single-dish lightcurve indicates that the outburst was associated with the central $\sim1$~pc core.  We found that this outburst accompanied the emergence of a new component, and the projected speed of this new component was $0.23c$ from 2007/297 (2007 October 24) to 2009/114 (2009 April 24).  We argue the site of $\gamma$-ray emission detected by \fermilat \ and jet kinematics in connection with $\gamma$-ray emission mechanism.}

\titlerunning{VLBI Monitoring of 3C~84}
\authorrunning{Nagai et al.}
   \maketitle
%

\section{Introduction}
NGC~1275 is the nearby Seyfert galaxy, the dominant member of Perseus cluster.  At the radio wavelengths, it is also known as the strong radio source 3C~84.  The radio structure of 3C~84 is complex, consisting of multiple lobe-like features (radio bubbles) on many different angular scales.  In the innermost region, there is a pair of lobes, which is probably formed by the jet activity originating in the 1959 outburst (e.g., Asada et al. 2006), extending to $\sim5$~pc from the core along north-south direction.  The northern counterjet becomes brighter and extends closer to the core with increasing frequency, suggesting free-free absorption by accretion disc rather than Doppler dimming (Romney et al. 1995).  Detection of northern counter jet infers the jet viewing angle of $30^{\circ}$-$55^{\circ}$ with the jet speed of $0.3$-$0.5c$ (Walker et al. 1994).  

The LAT of \fgrst \ found GeV $\gamma$-ray emission from NGC~1275 (Abdo et al. 2009, hereafter A09).  \fermi \ $\gamma$-ray flux is about 10 times higher than the upper limit constrained by EGRET.  This indicates that the source is variable on timescales of years to decades, and therefore the size of $\gamma$-ray emission region is likely to be smaller than $\sim10$~pc.  Interestingly, University of Michigan Radio Astronomy Observatory (UMRAO) revealed a radio flare starting in 2005.  MOJAVE observations also found radio brightening in the central $\sim1$~pc core between 2007 and 2008.  This coincidence of radio and $\gamma$-ray flares may indicate that the $\gamma$-ray emission is associated with the radio core.  Thanks to its proximity, 3C~84 is an excellent laboratory to study the jet physics very close to the core in connection with $\gamma$-ray emission mechanism.  We report here the results from data of VLBI and long-term single dish monitoring. 

\section{Observations and Data Reduction}
\subsection{Archival Data}
We used archival data of VLBI Exploration of Radio Astrometry (VERA) from 2006 May to 2008 May (2006/134, 2006/143, 2006/346, 2007/142, 2007/258, 2007/297, 2007/324, 2007/361, 2008/035, 2008/063, 2008/106, 2008/141\footnote{Observing sessions are denoted by year/(day of the year).}).  Observations were carried out with four VERA stations at 22.2~GHz, where 3C~84 was being used as a bandpass calibrator or a fringe finder.  Typically five scans of 5-minutes duration were obtained.  Left hand circular polarization (LHCP) was received and sampled with 2-bit quantization, and filtered using the VERA digital filter unit (Iguchi et al. 2005).  The data were recorded at a rate of 1024~Mbps, providing a bandwidth of 256~MHz in which 14 IF-channels per a total of 16 IF-channels of 16~MHz bandwidth were assigned to 3C~84.  For 2006/134, 2006/143, and 2007/297 data, only 1 IF-channel with 8~MHz bandwidth was assigned, and for 2007/142 data, 2 IF-channels with 8~MHz bandwidth were assigned.  Correlation processes were performed with the Mitaka FX correlator (Chikada et al. 1991).  

\subsection{New Observations with VERA and JVN}
The Japanese VLBI Network (JVN: Fujisawa 2008) observations were carried out on 2008/354 using the four VERA stations at 22.2~GHz, and on 2008/356 using the Yamaguchi 32-m telescope, the Tsukuba 32-m telescope, and the four VERA stations at 8.4~GHz.  The VERA observation was carried out on 2009/114 at 22.2~GHz.  Right hand circular polarization was received at 8.4~GHz, and LHCP was received at 22.2~GHz.  The data were recorded at the rate of 128~Mbps.  Data correlation was performed with the Mitaka FX correlator.  VERA observations were performed in dual-beam phase referencing mode, but in the present paper we report the analysis using the one-beam data.  

\subsection{Single-Dish Monitoring with Effelsberg}	
The flux density measurements at the Effelsberg 100-m telescope of the Max-Planck-Institut f\"{u}r Radioastronomie (MPIfR) were obtained during regular calibration and pointing observations at 22~GHz. The measurements were done using cross-scans in azimuth and elevation direction. The data reduction was done in the standard manner as described by Kraus et al. (2003).  The measured antenna temperatures were linked to the flux-density scale using primary calibrators like NGC~7027, 3C~286, and 3C~48 (Ott et al. 1994; Baars et al. 1977).

\subsection{VLBI Data Reduction}
Data reduction was performed using the NRAO Astronomical Imaging Processing System (AIPS).  A priori amplitude calibration and fringe fitting were performed by usual manner.  For JVN observation at 8.4~GHz, amplitude calibration was performed by scaling the correlated flux to the flux obtained with single dish observation.  For more on the detailed description, see Nagai et al. (2010). 

\section{Results}
\begin{figure}
   \centering
   \includegraphics[width=8cm]{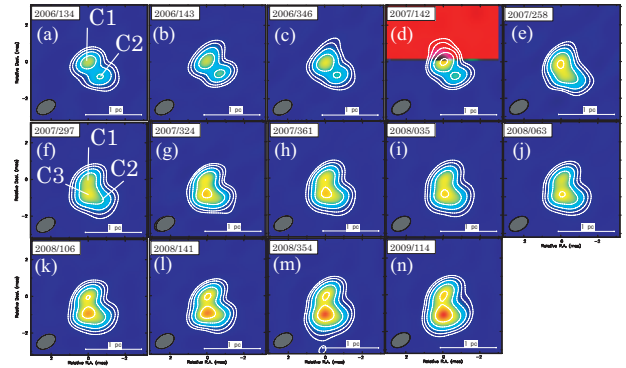}
      \caption{VERA images of 3C~84 at 22~GHz.  All images are shifted in reference to the northern component (component C1).  The contours are plotted at the level of 4.18, 8.36, 16.72, 33.44, and 66.88 \% of the peak intensity (4.989~Jy/beam) on 2009 April 24.  The restoring beam (1.1$\times$0.7~mas, position angle of $-60^{\circ}$) was set to make images uniform.}
         \label{fig1}
   \end{figure}

\begin{figure}
   \centering
   \includegraphics[width=7.5cm]{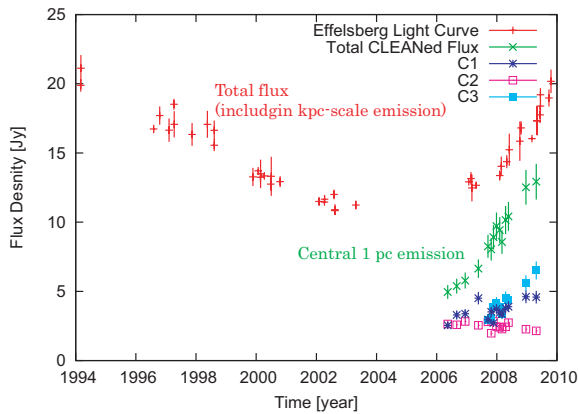}
      \caption{Radio lightcurve of 3C~84. Pluses: The Effelesberg light curve of 3C~84 at 22~GHz.  Crosses: total CLEANed flux of VERA observation at 22.2~GHz.  Asterisks:  The light curve of component C1.  Open squares: The light curve of component C2.  Filled squares: The light curve of component C3.}
         \label{fig2}
   \end{figure}

Figure \ref{fig1} shows total intensity images of the central $\sim1$~pc region at 22.2~GHz.  In the first 4 epochs, component C2 was visible at the position separated by $\sim1$~mas from the central core (C1) in a position angle of $-141^{\circ}$.  The alignment of these components was similar to that in the $\gamma$-ray quiet phase (Dhawan et al. 1998).  During first 4 epochs, component C2 showed no significant motion relative to component C1.  Remarkably, a new component (C3) suddenly emerged to the south of the central core on 2007/258 (Fig. \ref{fig2}(e)) despite only 4 months having passed since the previous observation (Fig. \ref{fig2}(d)).  Moreover, component C3 was clearly resolved on 2007/297 (Fig. \ref{fig2}(f)), separated by only one month from the previous epoch observation in which component C3 was blended with the other components.  In later epochs, component C3 appeared more significant.  

In Figure \ref{fig2}, we show the light curve monitored with the Effelsberg 100-m telescope and total CLEANed flux at 22.2~GHz with VERA as a function of time.  The tendency of flux increase in the central 1-pc core resembles to total flux increase. The difference between them, probably the emission from the extended feature, is almost constant ($\sim5$~Jy).  In order to test any change of the extended feature, we also analyzed 8-GHz data observed with JVN, which has better coverage of short interferometer baselines compared to 22-GHz data.  No significant change of the extended feature is confirmed compared to that in $\gamma$-ray quiet phase (see Nagai et al. 2010).  It is obvious that the flux increase starting in 2005 is originated in the central 1-pc core.  The flux density of each component is also plotted in Figure \ref{fig1}.  Component C3 showed the most significant increase in flux.

\begin{figure}
   \centering
   \includegraphics[width=7.5cm]{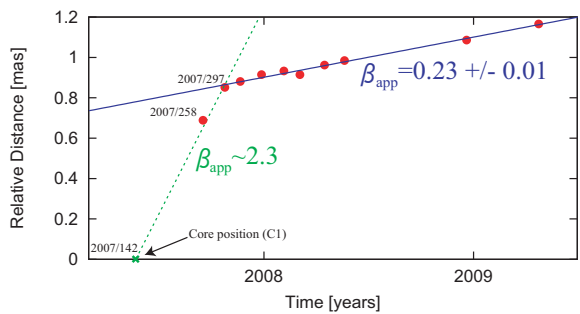}
      \caption{Plot of the separation between component C3 and component C1.  The error bar is smaller than the size of each symbol.  The blue solid line represents a linear fit to the data from 2007/297 to 2009/114.  The green broken line represents the one from 2007/142 to 2007/297, assuming that component C3 was ejected from the position of component C1 on 2007/142 (see \S4.2).}
         \label{fig3}
   \end{figure}

Figure \ref{fig3} shows the separation between the components C1 and C3 as a function of time.  The position of the components C1 and C3 was derived from the two-dimensional Gaussian fit in the interferometric ($u, v$)-plane using the ``modelfit" task in Difmap.  It is difficult to measure the positional error of each component quantitatively from the interferometric data in each epoch independently.  We thus employed the method described in Homan et al. (2001).  We initially set the uncertainty for each data point equal to unity, and then we performed a linear fit to the data assuming the motion with constant speed to obtain a preliminary $\chi^{2}$.  Taking this preliminary $\chi^{2}$, we then uniformly rescaled the uncertainty of each data point such that reduced-$\chi^{2}$ to be unity.  Finally, the positional error of each data point is estimated to be 0.013~mas.  This error is typically two times larger than the one estimated from the signal-to-noise ratio (SNR) such that $\theta_{\mathrm{beam}}/\mathrm{SNR}$, where $\theta_{\mathrm{beam}}$ is the beam size.  This fit results in an apparent speed of 0.20$\pm$0.01~mas/yr (projected speed of $0.23\pm0.01c$) towards the south.  This is approximately consistent with the jet speeds in $\gamma$-ray quiet phase (Dhawan et al. 1998).    The direction of movement of the new component differs from the alignment of the components C1 and C2 by $\sim40^{\circ}$ on the projected plane.  We note that we did not include the data on 2007/258 to this fit because component C3 might have moved faster before 2007/297 (see \S 4).

\section{Discussion}
In this section, we firstly discuss the intrinsic speed of the new component (C3), and then compare the jet speed expected from the SED modeling including GeV $\gamma$-ray data with the one measured using VLBI.  Finally, we argue the possible site of GeV $\gamma$-ray emission.
\subsection{Constraints on the Speed of C3}
\begin{figure}
   \centering
   \includegraphics[width=7.5cm]{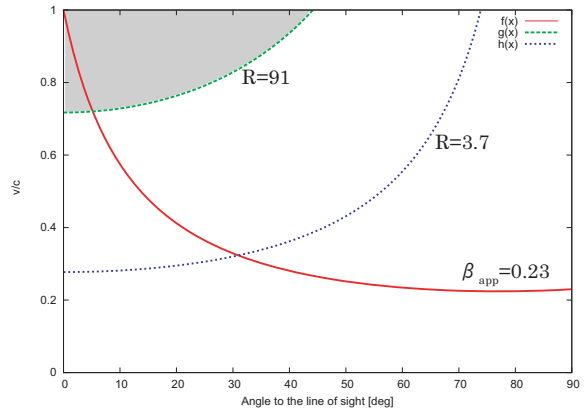}
      \caption{Constraints on the speed and the angle to the line of sight.  Red solid line represents $\beta=0.23$. Blue dotted line represents $R=3.7$.  Green broken line represents $R=91$.  Gray shaded region represents the condition required from non-detection of the counterjet.}
         \label{fig4}
   \end{figure}

Under the simple beaming model for a symmetric jet system with equal and constant speeds, apparent speed of approaching jet is 
\begin{equation}
\beta_{app}=\frac{\beta\sin{\theta}}{1-\beta\cos{\theta}}.
\end{equation}
where $\beta$ is the intrinsic jet speed and $\theta$ is the angle to the line of sight.  The ratio of brightness of the jet and counterjet components is 
\begin{equation}
R=\left(\frac{1+\beta\cos{\theta}}{1-\beta\cos{\theta}}\right)^{\eta}
\end{equation}
where $\eta$ can be $(2-\alpha)$ or $(3-\alpha)$, depending on a continuous jet or a single component ($\alpha$ is the spectral index; $S\propto\nu^{-\alpha}$).  In Figure \ref{fig4}, the curve for $\beta_{app}=0.23$ is represented by red solid line and the region for $R>91$ is represented by the gray-shaded region.  The region for $R>91$ is the lower limit constrained by non-detection of counterjet in Figure \ref{fig1}.  Here we assumed that a single component with $\alpha=0.5$.  For these two conditions, the jet speed and the angle to the line of sight are about $\beta>\sim0.72$ and $\theta<\sim5^{\circ}$.  This constraint on the jet viewing angle is much smaller than the angle to the line of sight estimated for $\sim10$~mas scale jets ($30$-$55^{\circ}$; Walker, Romney, \& Vermeulen 1994).  Such a change in angle can occur if the jet is precessing.  The change in the jet direction in Figure \ref{fig1} might be also explained by the precession. 

A very small angle to the line of sight is not required under the free-free absorption model (e.g., Walker, Romney, \& Vermeulen 1994).  The pc-scale northern counterjet has a strongly inverted spectral index, which is consistent with free-free absorption by ionized material along the line of sight to the counterjet.  The spectral fits to the multi-frequency VLBI data showed that the opacity increases closer to the core (Walker et al. 2000).  The new jet components originated in the recent outburst still reside very close to the core ($\sim1$~pc), and therefore the counterjet component must suffer from very strong absorption.  The brightness ratio between the jet and counterjet can be smaller if we remove the absorption effect.  If we adopt the brightness ratio between the jet and counterjet in 10~mas scale ($R=3.7$; Walker et al. 2000), the jet speed and the angle to the line of sight are about $0.3c$ and $\theta=30^{\circ}$. 

\subsection{Discrepancies of jet speeds}
Modeling of the broad-band SED including \fermi \ data in a framework of blazar scenario with moderate beaming requires the bulk Lorentz factor of $\Gamma=1.8$ ($\beta=0.83$) and the viewing angle ($\theta$) of $25^{\circ}$ (A09).  Our observed apparent speed of $\beta_{app}=0.23$ would imply that $\beta=0.83$ can be achieved for $\theta<\sim3^{\circ}$.  This disagrees with not only the jet viewing angle derived from the SED modeling but also $R=3.7$ derived from jet-counterjet ratio in 10~mas scale.  If our non-detection of the counterjet is solely owing to the relativistic beaming ($R>91$), $\beta=0.83$ and $\theta=25^{\circ}$ can be reproduced.  However, this is not consistent with our observed apparent motion ($\beta_{app}=0.23$).  In any case, under the simple beaming model, observed jet speed and jet-counterjet ratio is not consistent with the jet speed and viewing angle derived from the SED modeling.     

Similar apparent slow-moving jet in spite of strong $\gamma$-ray emission is seen in some BL Lac objects and a radio galaxy M~87 (e.g., Giroletti et al. 2004; Edwards \& Piner 2002; Kovalev et al. 2007).  A spine sheath model has been suggested to ease this problem: a fast spine jet produces the strong TeV emission by inverse Compton scattering of the radio photon from a surrounding slow layer (Ghisellini et al. 2005).  Limb-brightening structure in Mrk~501 and M~87 is consistent with the spine-sheath model (e.g., Giroletti et al. 2008; Kovalev et al. 2007).  However, it is difficult to test limb-brightening in 3C~84 even with VLBA observation at 43~GHz (Dhawan et al. 1998), due to lack of resolution.  Upcoming higher resolution telescope VSOP-2 (Tsuboi et al. 2009) would be valuable in this respect.  

Decelerating jet model is also possible to solve this problem (Georganopoulos \& Kazanas 2003).  We note that there is one data point which does not fit into the regression line of $0.23c$ in Figure \ref{fig3}.  This could be a signature of the deceleration.  However, we need further careful study due to lack of resolution and possible absorption effect (see Nagai et al. (2010)).  

\subsection{Implication for the site of $\gamma$-ray emission}
It is of great interest whether $\gamma$-ray emissions come from the close vicinity of the core or the distant component from the core in AGNs.  For example, in M~87 which is another low power (FRI) radio galaxy known as $\gamma$-ray source, there is a prominent component separated from $\sim100$~pc from the central core, so-called HST-1.  In 2005, significant flux increase was observed at radio and X-ray bands in HST-1 (Cheung et al. 2007).  During this flare, TeV excess was also confirmed, allowing the speculation that HST-1 might be the site of $\gamma$-ray flare.  In contrast to the 2005 flare, a TeV flare in 2008 was accompanied by a strong increase of the radio flux from the core whereas no obvious radio brightening in HST-1 (Acciari et al. 2009).  

Due to the lack of $\gamma$-ray data between EGRET era and \fermi \ era, we do not know when the GeV $\gamma$-ray flare was started exactly in NGC~1275.  However, it is clear that $\gamma$-ray activity of NGC~1275 is now in active phase compared to EGRET era, and this $\gamma$-ray variability is very similar to the radio flux increase (A09).  Moreover, the radio flux increase is likely to be originated in the central 1-pc core.  This allows us to expect that the $\gamma$-ray emission might be associated within the central 1-pc core.

Although there is a long-term correlation between radio and $\gamma$-ray lightcurves, the problem of no exact correlation between $\gamma$-ray and radio during a timescale within a year remains; while radio flux increases continuously, the $\gamma$-ray emission shows no significant time variation except a small flare in April-May 2009.  There is room for further investigation in this respect.


%
%

\begin{acknowledgements}
We are grateful to the staffs of all the VERA and JVN stations for their assistance in observations.  This work is based on observations with the 100-m telescope of the MPIfR at Effelsberg.  We thank Alex Kraus for providing us with additional archival Effelsberg data (1994-2004).

\end{acknowledgements}

\end{document}